\documentclass{article}

\usepackage[utf8]{inputenc}
\usepackage[letterpaper]{geometry} 
\usepackage{fancyhdr} 
\usepackage{lastpage} 
\usepackage{color,soul} 
\usepackage{xcolor}
\usepackage{graphicx}	
\usepackage{float}
\usepackage[colorlinks=true,linkcolor=blue,citecolor=blue,urlcolor=blue]{hyperref}%
\usepackage{subcaption}

\usepackage[round]{natbib}
\title{\textbf{Observing M Dwarfs UV and optical flares from a CubeSat and
their implications for exoplanets habitability}}
\author{\href{https://orcid.org/0000-0003-4064-4268}{{\textbf{Julien Poyatos}}}$^{1*}$
, \href{https://orcid.org/0000-0002-4227-9308}{{\textbf{Octavi Fors}}}$^{1}$
, \href{https://orcid.org/0000-0003-0173-5888}{{\textbf{José Maria Gómez Cama}}}$^{2}$
}

\date{%
 $^1$Departament de F{\'i}sica Qu{\`a}ntica i Astrof{\'i}sica, Institut de Ci{\`e}ncies del Cosmos (ICCUB), Universitat de Barcelona, IEEC-UB, Mart\'{\i} i Franqu{\`e}s 1, E-08028 Barcelona, Spain\\%
 $^2$Departament d'Enginyeria Electr{\`o}nica i Biom{\`e}dica, Institut de Ci{\`e}ncies del Cosmos (ICCUB), Universitat de Barcelona, IEEC-UB, Mart\'{\i} i Franqu{\`e}s 1, E-08028 Barcelona, Spain\\
 $^*$Corresponding author\\[2ex]%
 }

\begin{document}

\begin{center}
IAC-22-B4
\end{center}
{\let\newpage\relax\maketitle}
\thispagestyle{fancy}
\begin{center}
    \section*{Abstract}
\end{center}
\par M dwarfs show the highest rocky planet occurrence among all spectral types, in some instances within the Habitable Zone. Because some of them are very active stars, they are often subject to frequent and powerful flaring, which can be a double-edged sword in regard of exoplanet habitability. On one hand, the increased flux during flare events can trigger the chemical reactions that are necessary to build the basis of prebiotic chemistry. On the other hand, sufficiently strong flares may erode exoplanets’ atmospheres and reduce their UV protection. Recent observations of flares have shown that the flaring flux can be $\times$100 times stronger in UV than in the optical. UV is also preferable to constrain more accurately both the prebiotic abiogenesis and the atmospheric erosion. For these reasons, we are developing a CubeSat payload concept to complement current flare surveys operating in the optical. This CubeSat will observe a high number of flaring M dwarfs, following an all-sky scanning law coverage, both in the UV and the optical to better understand the different effective temperatures as wavelengths and flaring status go. This will complement the bright optical flares data acquired from the current ground-based, high-cadence, wide FoV surveys. Another scientific planned goal is to conduct few-minute after-the-flare follow-up optical ground-based time-resolved spectroscopy, that will be triggered by the detection of UV flares in space on board of the proposed CubeSat. Finally, the study of M dwarfs stellar activity in the UV band will provide useful data for larger forthcoming missions that will survey exoplanets, such as PLATO, ARIEL, HabEx and LUVOIR. 
\bigbreak
\noindent \textbf{Keywords:} M dwarf, flares, CubeSat, exoplanets, habitability, UV

\twocolumn
\section{Introduction}
\subsection{Stellar flares and exoplanet habitability}

M dwarfs make up to $\sim$75\% of the stellar population and frequently host terrestrial planets in their habitable zones \citep{henry2006solar,covey2008luminosity}. Their small radii and low temperatures enable the detection and atmospheric characterization of these planets, making M dwarfs excellent targets in the search of habitable exoplanets \citep{kaltenegger2009transits}. However, these stars are known to be very active (with stars later than M4 being fully convective), making them prone to frequently emit large stellar flares throughout their lifetimes \citep{kowalski2010white}. Powerful flares are also often associated with Coronal Mass Ejections (CMEs), streams of charged particles ejected into space. Flares and CMEs can play an important role in planetary evolution and habitability, mainly through atmospheric erosion, and act as stressors for surface life \citep{tarter2007reappraisal}. 
The span of the HZ around M dwarfs is much smaller than that around earlier-type stars. In consequence, exoplanets orbiting in the HZ of M dwarfs are more affected by flares and CMEs. In the most extreme scenario, intense flaring activity could render the surface of an exoplanet uninhabitable, leaving only a chance for life to develop in an ocean or underground \citep{tilley2019modeling,estrela2018superflare}. More optimistically, flares may catalyse biosignatures, or serve as a source for otherwise scarce visible-light photosynthesis on planets orbiting M dwarfs. Indeed, recent work suggests that sufficient UV radiation might trigger the synthesis of prebiotic chemistry, turning inorganic matter on a planet's surface (mineral dust, meteoritic dust, etc) into the bases of life (sugars, nucleobases, carboxylic acids, amino acids, etc) \citep{ranjan2017surface,rimmer2018origin}. The typical effective temperatures of M dwarfs makes them unlikely to produce sufficient UV radiation for this purpose, in this case the required amount of UV radiation can only come from flares \citep{chen2021persistence}. In sum, there seems to exist a "sweet spot" where M dwarfs could be flaring just enough to provide enough UV radiation for the trigger of prebiotic chemistry, but not too much to induce atmospheric erosion and surface sterilization. It is therefore essential to gather information on the frequency and energy of host star's flares to better characterise the habitability of their surrounding planets.

\subsection{CubeSats for astrophysics}

CubeSats are a type of small satellites (i.e. less than 100kg) composed of one or several standardised units of 10x10x10cm. This format allows them to use off-the-shelf components, drastically lowering the cost and construction time of satellite missions \citep{poghosyan2017cubesat}. These are getting more and more popular in the recent years, allowing individuals, startups of universities to lead their own missions. Space agencies are also adopting CubeSats through their New Space strategies. 

In this context, we are studying a CubeSat concept to implement in a future mission, in order to observe M dwarfs flares in the UV and the optical. This $\geq$6U CubeSat will gather light curves and images of flaring M dwarfs in both bands split in several filters to compare the difference in flux ratios. These data products will allow to complement our understanding on stellar physics for future missions (such as PLATO, ARIEL, HabEx and LUVOIR), and to provide new inputs in the Star-Planet Interaction models. The CubeSat will also be used to detect flares as early as possible in UV, then trigger alerts to ground-based observatories to observe few-mins after-the-flare time-resolved follow-up spectroscopy. 
In general, UV flares reach their peak faster than optical flares. We could, in the best case, use the detection of UV flares as a trigger for optical flares spectroscopy, which would give precious data on the flare physics scenario (mainly the evolution of the continuum of H$_{\alpha}$ and optical lines associated with CMEs) \citep{fuhrmeister2022high,wu2022broadening}. Finally, it will be able to detect Quasi-Periodic Pulsations (QPPs) \citep{fleming2022new,doyle2022doubling}. These flares happening at an almost fixed period are interpreted as a modulation of the flare's chromospheric thermal emission through periodic triggering of reconnection by external magnetohydrodynamic oscillations in the corona \citep{doyle2022doubling,fleming2022new}. Observations of multiple flares per active stars will allow to test this interpretation. 

We report in this paper the first calculations on the expected number of detectable flares per day by our CubeSat concept, for different target types. This work will be organised as follow: In Section \ref{section:mat_met}, we present the catalog, equations and component characteristics we used for the calculations. In Section \ref{section:results}, we provide the calculation results for different types of M dwarfs, according to their mass and distance. In Section \ref{section:discussion}, we discuss possible consequences on the CubeSat mission primary and secondary science goals and possible future mission prospects. We conclude in Section \ref{section:conclussion}.

\section{Materials and methods}
\label{section:mat_met}

\subsection{The TOI catalog}

TESS (Transiting Exoplanet Survey Satellite) is searching the entire sky since 2018 for transiting exoplanets \citep{ricker2014}. Since then, more than 5000 TESS Objects of Interest (TOIs), have been identified. These includes both new planet candidates found by TESS and previously-known planets recovered by TESS observations. \cite{howard2022flaring} explored the light curves of the 2250 non-retired TOIs from the Sectors 1 \& 2 and extrapolated the flare rates through flare injection and recovery for their 2096 host stars. 

Flare emission from stars is described by a power law in which higher energy flares are emitted less frequently than lower energy flares. Flare Frequency Distribution (FFDs) are computed for each TOI host star given the cumulative rate at which flares of energy $E$ or larger are observed per day and the total observing time. FFDs are fit in a log-log space in the usual way:
\begin{equation}
    \log \nu = (1 - \alpha) \log E + \beta
    \label{eq:ffd}
\end{equation}
where $\nu$ is the number of flares with an energy greater than or equal to $E$ erg per day, $1-\alpha$ describes the frequency at which flares of various energies occur, and $\beta$ determines the overall rate of flaring. This catalog places upper limits on the flare rate of host stars. Injection and recovery tests of flares are performed for non-flaring stars which enables to place upper limits on their flare rates when combined with the total observation time. 

\begin{figure}
    \centering
    \includegraphics[clip,trim={0 0 0 0},width=0.48\textwidth]{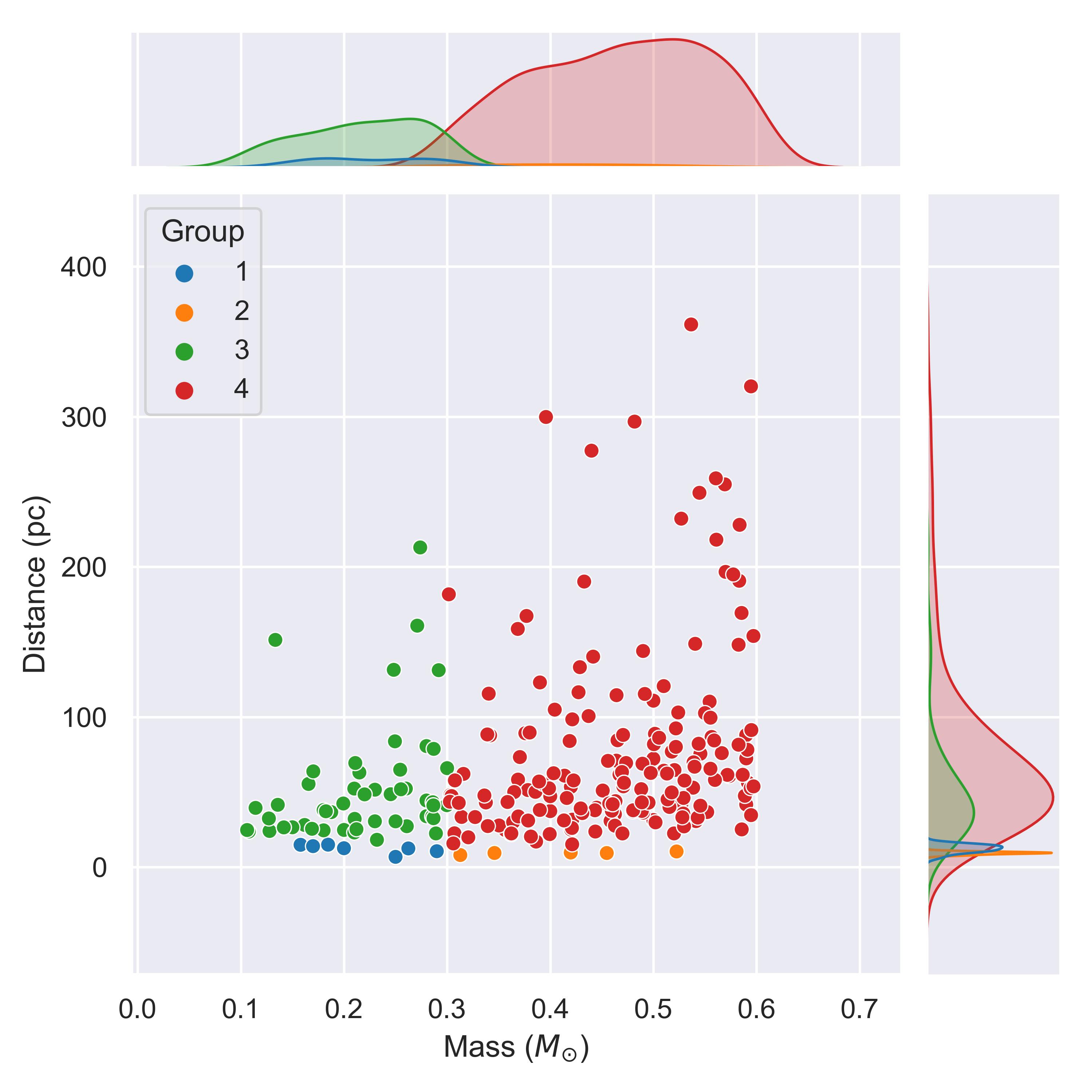}
    \caption{Mass/distance diagram of the targets from the first two groups. Top and right histogram represent the distribution of the targets by mass and distance, respectively.}
    \label{fig:mass/distance}
\end{figure}

We cross-match M dwarfs from the TOI catalog with the TIC catalog to obtain their distance and radius. Following the method in \cite{howard2022flaring}, we separate the stars in different groups according to their mass and distance in order to facilitate comparisons. The different groups are sorted as in Table \ref{tab:groups}. \cite{winters2021volume} determines 15 pc as the demarcation distance between nearby and distant M-dwarfs. We extrapolate an equivalent demarcation distance for larger mass groups such that the apparent magnitude of stars in each group is approximately comparable to M dwarfs observed from 15 pc. For this work, we only consider in Section \ref{section:number_of_flares} flares from M dwarfs (corresponding to the groups 1-4 in Table \ref{tab:groups}). However, our CubeSat concept could easily detect K dwarf flares as well (corresponding to groups 5-8 in Table \ref{tab:groups}). We plot in Figure \ref{fig:mass/distance} the distribution of the targets from Groups 1-4 on a mass and distance diagram. 

\begin{table*}
    \centering
    \begin{tabular}{c|c|c|c|c}
    \hline
      Group number  & Spectral type & Mass [$M_\odot$] & Distance [pc] &  Number of targets\\
      \hline
      1 & Late M & 0.1 - 0.3 & 0 - 15 & 10 \\
      2 & Early M & 0.3 - 0.6 & 0 - 15 & 5 \\
      3 & Late M & 0.1 - 0.3 & 15 - 36 & 53 \\
      4 & Early M & 0.3 - 0.6 & 15 - 36 & 210 \\
      5 & Late K & 0.6 - 0.75 & 0 - 36 & 20 \\
      6 & Early K & 0.75 - 0.9 & 0 - 36 & 13 \\
      7 & Late K & 0.6 - 0.75 & 36 - 80 & 160 \\
      8 & Early K & 0.75 - 0.9 & 36 - 80 & 235 \\
      9 & Late G & 0.9 - 0.98 & 0 - 80 & 33 \\
      10 & Early G & 0.98 - 1.06 & 0 - 80 & 301 \\
      11 & Late G & 0.9 - 0.98 & $>$ 80 & 203 \\
      12 & Early G & 0.98 - 1.06 & $>$ 80 & 38 \\
      \hline
    \end{tabular}
    \caption{TOI host stars are divided into 12 different groups according to their mass and distance. From left to right: number of the group, corresponding spectral type, mass range, distance range (in pc), and number of targets within each group. Out of the 2096 target stars, 968 do not fall in any groups and are therefore not considered.}
    \label{tab:groups}
\end{table*}

\subsection{Component characteristics}

Since one of this CubeSat's scientific goal is to observe flares in two different bands (UV and optical), it needs to embark either two sensors observing each in one band, or one sensor observing in both bands simultaneously. In this work we chose to explore the first option. We used for the following ID4MTVISB-CL\footnote{\url{https://en.idule.jp/wp/wp-content/uploads/2020/07/ID4MTVISB-CL_E_20200715.pdf}}, a commercial CMOS manufactured by Gpixel. This backside illuminated CMOS image sensor provides high sensitivity, low noise, and high dynamic range in both UV and optical, making it adequate for the scientific purpose of this mission. We give in Table \ref{tab:cmos} a few key technical characteristics of this CMOS, mainly the ones that were used for the calculations. We selected the two wavelength ranges as follows: UV = [200 nm - 275nm], Optical = [378nm, 555nm]. This UV band contains emission lines from transition and chromospheric regions (C IV, He II and Mg II) \citep{fuhrmeister2022high}, and the optical band contains H$_{\alpha}$ line profiles \citep{wu2022broadening}. Data from these emission lines will be extremely useful to study short- and long-term variability of active M dwarfs, to provide crucial inputs to stellar models and to understand how flare color ratios change over flare duration, effective temperature, stellar mass, stellar rotation period, and stellar age. The optical band contains the exo-auroral emission lines, increasing the star/planet contrast by orders of magnitude.

\begin{figure*}
    \vspace*{-5mm}
    \centering
    \includegraphics[clip,trim={0 0 0 0},width=0.75\textwidth]{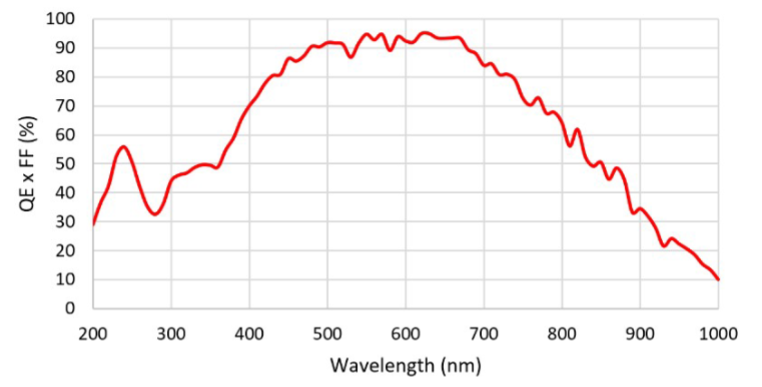}
    \caption{The quantum efficiency curve of the ID4MTVISB-CL CMOS image sensor.}
    \label{fig:QE_sensor}
\end{figure*}

We average the quantum efficiency for each band from the spectral response curve provided by the manufacturer (Figure \ref{fig:QE_sensor}). This gives us QE$_{\rm UV} \approx 0.45$ and QE$_{\rm optical} \approx 0.90$. We detail in Table \ref{tab:cmos} the characteristics that will be used for the calculations in Section \ref{section:number_of_flares}.

For the telescope, we considered a 9cm aperture-class telescope, similarly to other CubeSats in construction with similar goals \citep{ardila2018star}.

\begin{table*}
    \centering
    \begin{tabular}{c|c|c|c}
    \hline
    Name & Meaning & Value & Units \\
    \hline
    $n_{\rm pix}$ & Effective pixel number & $2048(H) \times 2048(V)$ & . \\
    $s_{\rm pix}$ & Pixel size & 11$\mu m \times 11 \mu m$ & $\mu m^2$ \\
    $Q_{\rm E-UV}$ & Average quantum efficiency in UV & 0.45 & .\\
    $Q_{\rm E-opt}$ & Average quantum efficiency in optical & 0.90 & .\\
    $RON$ & Read-out-noise & 0.2 & $e^- /pixel$ \\
    $DARK$ & Dark current & 1.6 & $e^- /s/pixel$ \\
    \hline
    \end{tabular}
    \caption{CMOS sensor opto-electrical characteristics.}
    \label{tab:cmos}
\end{table*}

\subsection{Expected number of observable flares per day}
\label{section:number_of_flares}

The bolometric energy released by an M dwarf flare is usually estimated by considering a 9000K blackbody curve \citep{allred2015unified}. However, this method is generally used to estimate the energy produced by very strong flare events by extrapolating their blackbody in the optical. \cite{2022arXiv220813674A} showed that this process is prone to large errors if the optical data is not combined to UV data, because optical bands are mostly sampling the Rayleigh-Jeans tail of the spectrum at high temperatures. The objective of this work is opposite: we want to determine what bolometric energy would produce a high enough SNR on the CubeSat's sensor. We therefore first re-estimate the temperature of the weakest recovered flare with the inverse of the Stefan-Boltzmann law:
\begin{equation}
    T = \left(\frac{R_e}{\sigma}\right)^{1/4} = \left( \frac{\partial \phi_e}{\partial A} \frac{1}{\sigma} \right)^{1/4} = \left(\frac{E_{bol}}{A} \frac{1}{\sigma}\right)^{1/4}
\end{equation}
where $R_e$ is the radiant exitance, $\sigma$ is the Stefan-Boltzmann constant, $\phi_e$ is the radiant flux emitted, $A$ is the surface of the emitting star and $E_{bol}$ is the bolometric energy of the weakest flare recovered. 

We then use the Planck's law to compute the blackbody radiation corresponding to this temperature \citep{kramm2009planck}: 
\begin{equation}
    L = \frac{2 \pi^4 k^4}{15h^3 c^2} T^4 \qquad W.m^{-2}.sr^{-1}
    \label{eq:planck}
\end{equation}
where $L$ is the spectral radiance, $h$ is the Planck's constant, $c$ is the speed of light, and $k$ is the Boltzmann's constant. 
We also derive from Equation ~\ref{eq:planck} the total photon irradiance: 
\begin{equation}
    L^p = \frac{4 \zeta(3) k^3}{h^3 c^2} T^3 \qquad photon.s^{-1}.m^{-2}.sr^{-1}
\end{equation}
where $\zeta$ is the Riemann zeta function and $\zeta(3) \approx 1.202056903159594$ (also known as Apéry's constant).

\begin{figure*}
    \centering
    \includegraphics[clip,trim={0 0 0 0},width=0.75\textwidth]{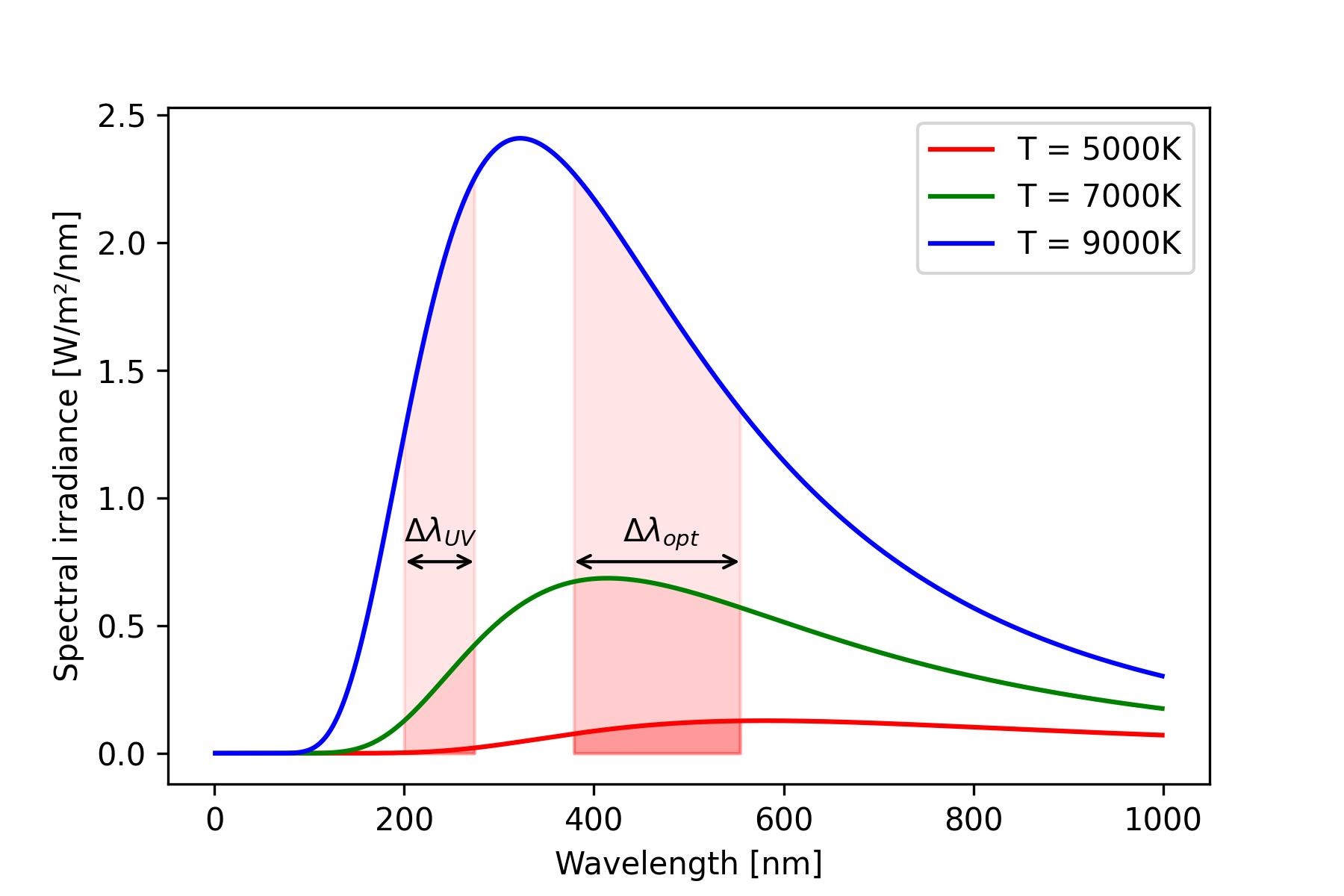}
    \caption{Spectral irradiance curves for black-bodies with temperature of 5000K (red), 7000K (green) and 9000K (blue). We color in shaded red the two bandwidth considered in this study.}
    \label{fig:blackbodyy}
\end{figure*}

Naturally, an increase in released bolometric energy (i.e. a flare) will be translated by an increase in temperature. This will impact the blackbody radiation by shifting the peak of the emission curve to shorter wavelengths. As a result, the amount of UV radiation released increases as the bolometric energy of the flare increases. 
In Figure \ref{fig:blackbodyy}, we plot three examples of blackbody radiations for different temperatures and highlight the two wavelength ranges we consider in this paper to highlight this effect. Even though the proportion of UV radiation is higher than optical radiation for high energy flares, the width of the UV band is smaller than that of the optical band. This will result in a lower SNR in UV compared to optical since less photons can be captured. 

To compute the in-band spectral radiance $F_e$ (and by extension the in-band spectral photon irradiance), we integrate the Planck's function over a finite range as follows: 
\begin{equation}
    F_e = \int_{\lambda_1}^{\lambda_2} L_{\lambda} d\lambda = \int_{\lambda_1}^{\infty} L_{\lambda} d\lambda - \int_{\lambda_2}^{\infty} L_{\lambda} d\lambda \qquad 
\end{equation}

and we use the code provided in \cite{widger1976integration} to estimate the two integrals. 
We then use the inverse square law to compute the number of photons received ($F_r$) at the CubeSat's entrance pupil, assuming each star as a perfect sphere emitting equally in every direction:
\begin{equation}
    F_r = \frac{F_e}{4\pi d^2} \qquad W.m^{-2}.nm
\end{equation}
where $d$ is the distance between the target star and the CubeSat (approximated as the distance between the target star and the Earth).

We use the ESO ETC textbook formula\footnote{\url{https://www.eso.org/observing/etc/doc/formulabook/node6.html}} to compute the number of photons emitted entering the CubeSat's optics, similar to the one used by \cite{gill2022low} :
\begin{equation}
    N = \frac{F_r . \Delta_{\lambda} . t . Q_E . S}{P}
    \label{eq:n_photons}
\end{equation}
where $N$ is the number of photons down-converted by the sensor, $F_r$ is the received flux of photons in $W.m^{-2}.nm$, $\Delta_{\lambda}$ is the wavelength range considered in $nm$, $t$ is the total exposure time in seconds, $Q_E$ is the average quantum efficiency of the band, S is the telescope surface in $m^2$ and $P$ is the energy of one photon. We consider the solid angle created by the length and width of the sensor as negligible.

Finally, the SNR is calculated with the formula given by the ESO ETC as: 
\begin{equation}
    SNR = \frac{N}{\sqrt{N + n_{pix} . RON^2 + n_{pix} . DARK . t}}
    ~\label{eq:snr}
\end{equation}
where $n_{pix}$ is the number of pixels of the sensor, RON is the read-out-noise in $e^- /pixel$, $DARK$ is the dark current in $e^- /s/pixel$ and $t$ is the exposure time.  

The CubeSat will operate following an all-sky scanning law and a heliosynchronous orbit, with a revisit time of 2 days. We consider an exposure time of $t = 90s$ and a processing time of 30s, meaning that a science image will be obtained every 2 minutes. The rest of the values used in Equation \ref{eq:n_photons} and Equation \ref{eq:snr} can be found in Table \ref{tab:cmos}.

\section{Results}
\label{section:results}
\begin{figure*}
    \centering
    \includegraphics[width=0.82\textwidth]{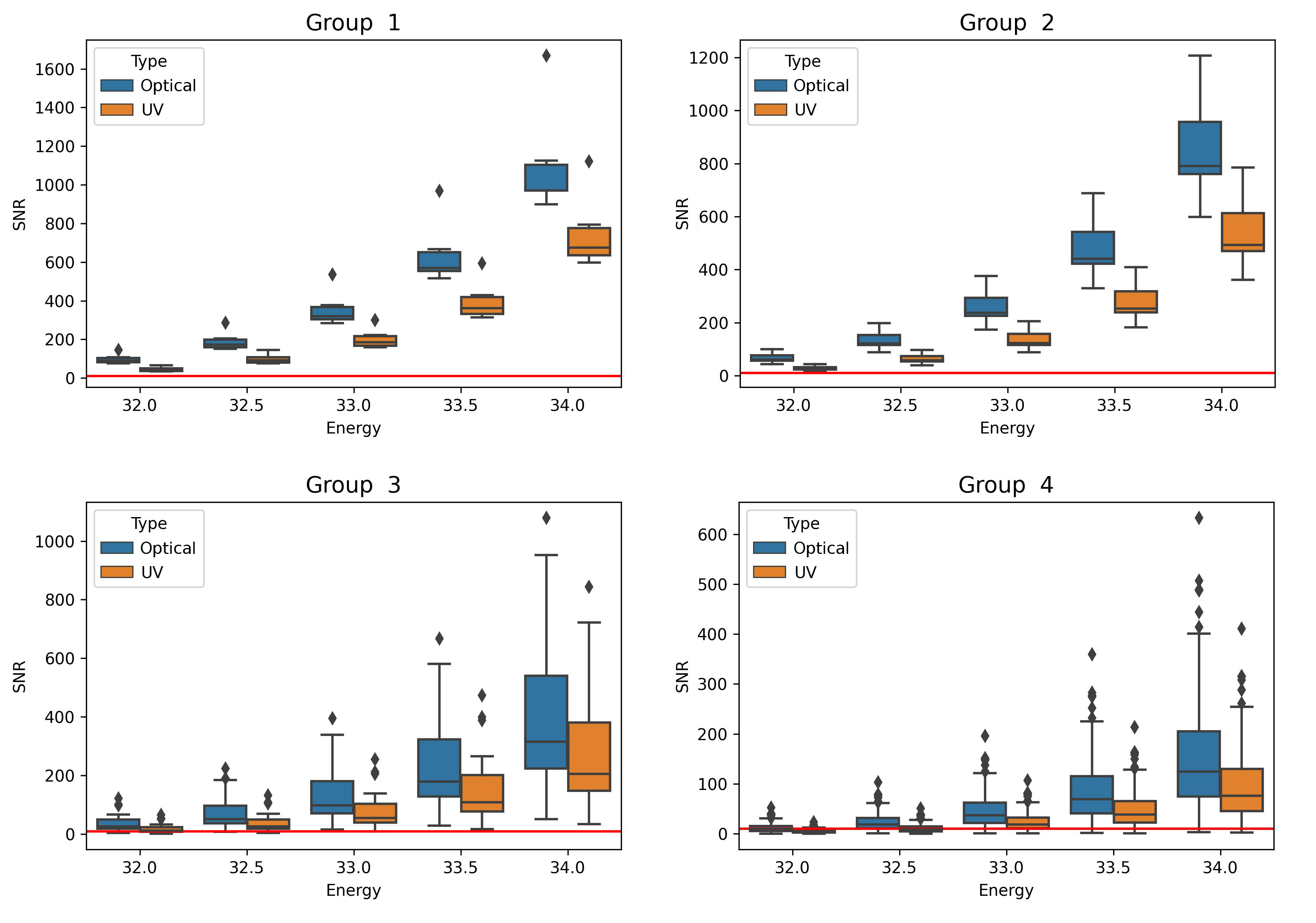}
    \caption{SNR corresponding to different flare energies for each target in the first four groups, for our optical band (blue) and UV band (orange). Horizontal bars correspond to, from bottom to top, statistical minimum, first quartile, median, third quartile, and statistical maximum. Outliers are represented by black diamonds. The red horizontal line corresponds to the value of SNR = 10.}
    \label{fig:boxplots}
\end{figure*}

The bolometric energy of each flare in the TOI catalog corresponds to the minimum energy for a flare to be observable and not blended in noise. However, the noise distribution and optical characteristics for our CubeSat will be different than those for TESS. Thus, we have to be careful when considering the weakest detectable flare. To address this we fix a threshold on the SNR. For a flare to be considered as detectable by our CubeSat, it has to produce a SNR $>$ 10 dB. We compute the SNR corresponding to different flare energies for each target in Groups 1-4, increasing the energy by $10^{0.5}$ erg intervals. We plot the results in Figure ~\ref{fig:boxplots}. All flares from Group 1 give an SNR $>$ 10 and are therefore considered as observable. However, this group contains drastically less targets as the three others, as can be seen by the small span of the box-plots. Results from Group 1 might be influenced by a poor-statistics bias and should therefore be considered with caution.

\begin{figure*}
    \centering
    \includegraphics[clip,trim={0 0 0 0},width=0.55\textwidth]{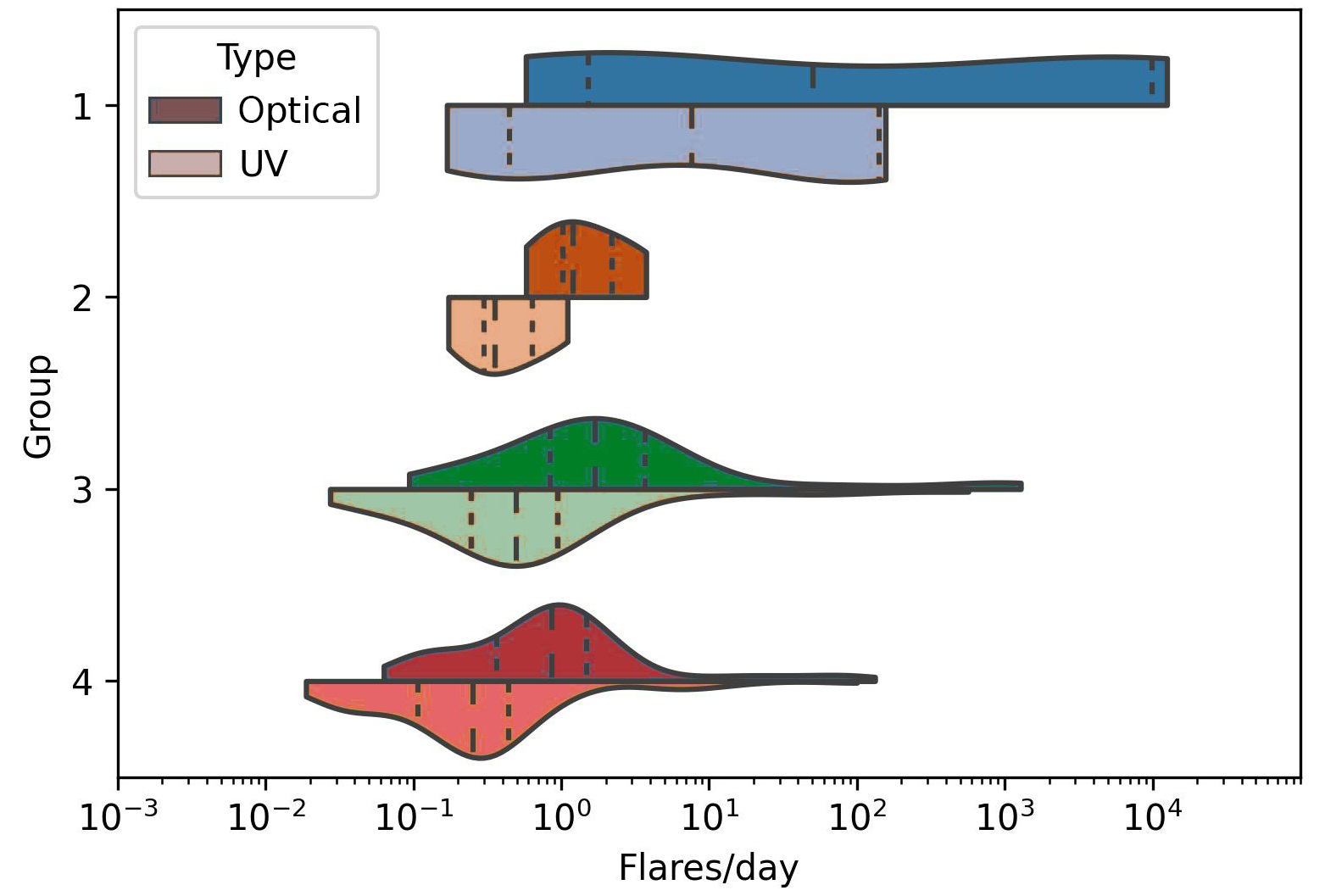}
    \label{fig:blackbody}
    \caption{Number of flares/day observable with a SNR $>$ 10. Vertical bars correspond, from left to right, to statistical minimum, first quartile, median, third quartile, and statistical minimum. The thickness of the box corresponds to the probability density (i.e. the box is thicker where there is a lot of data points). Dark shaded boxes corresponds to values for our optical band, and light shaded regions corresponds to values for our UV band. For clarity, the color of each group is similar to the one used is Figure \ref{fig:mass/distance}.}
    \label{fig:violin}
\end{figure*}

The amount of flares of a given energy over a period of time is determined by the FFD, as described in Equation \ref{eq:ffd}. \cite{howard2022flaring} estimates the $\alpha$ and $\beta$ parameters of each target in the TOI catalog through 200 flare injection and recovery tests. We then compute the FFD of each target from Groups 1-4, then crossmatch the data with the results from Figure ~\ref{fig:boxplots}. This gives us the number of flares per day with enough energy to be observed with a SNR $>$ 10, for each target. We plot the result in Figure ~\ref{fig:violin}. 

\section{Discussion}
\label{section:discussion}
Since Groups 3 and 4 represent nearly 95\% of the targets in the first four groups, we conclude that this set of components would allow our CubeSat to observe $\sim 10^{-1} - 10^{1}$ flares/day. Nearby bright targets from Group 1 drastically improve this value, with some specific targets producing nearly $10^4$ observable flares/day. Assuming the distribution of M dwarfs in the entire sky is the same as in the TESS Sectors 1 \& 2 and considering the CubeSat's revisit period of 2 days, we would gather enough UV and optical flare observations for the realization of the scientific goal in a $\sim$ 1-2 year mission lifetime. However, the CubeSat's optics have not been confirmed yet, so we advocate to reproduce similar calculations once the final CMOS and telescope are chosen. 

\section{Conclusion}
\label{section:conclussion}
In this paper, we used the TOI catalog complemented by injected flares from \cite{howard2022flaring} to determine the rate of M dwarfs flares detectable by our CubeSat concept. We used for the calculations two components likely to be used for this mission: the ID4MTVISB-CL UV and optical imaging CMOS sensor from Gpixel and a 9cm aperture-class telescope. We separated M dwarfs in the catalog according to their mass and distance, before determining the SNR of the weakest detectable flare for each target. We then used the $\alpha$ and $\beta$ parameters provided by \cite{howard2022flaring} to determine the FFD of each target. We cross-matched the flare's SNR values and the FFDs to determine the number of flares detectable per day by our CubeSat with an SNR $>$ 10. We conclude that a mission lifetime of $\sim$ 1 - 2 years should provide enough UV and optical flare observations to accomplish the described scientific goals. 

\section*{Acknowledgements}
This work made use of NASA ADS Bibliographic Services. This research has made use of several public tools developed at CDS, Strasbourg Observatory, France. OF and JMC acknowledge the support by the Spanish Ministerio de Ciencia e Innovaci\'{o}n (MICINN) under grants PID2019-105510GB-C31 and PGC2018-098153-B-C33, respectively. OF and JMC acknowledge the support through the ``Center of Excellence Mar\'{i}a de Maeztu 2020-2023'' award to the ICCUB (CEX2019-000918-M). This work made use of Astropy, a community-developed core Python package for Astronomy, and the NumPy, SciPy, and Matplotlib Python modules.

\small \bibliography{citation}
\end{document}